\documentclass[a4paper]{spie}

\usepackage[sc, hang]{caption}
\usepackage{graphicx}
\usepackage{graphics}
\usepackage{makeidx}
\usepackage[pass]{geometry}
\usepackage{booktabs}
\usepackage{marvosym}
\usepackage[hang]{subfigure}
\usepackage{SIunits}
\usepackage{rotating}
\usepackage{csquotes}
\usepackage{url}
\usepackage{multicol}
\usepackage{multirow}
\usepackage{overpic}
\usepackage{amsmath}
\usepackage{commath}
\usepackage{bm}
\usepackage{rotating}
\usepackage{psfrag}
\usepackage{parallel}
\usepackage{fancyhdr}
\usepackage[breaklinks]{hyperref}

\usepackage[english]{babel} 
\usepackage{eso-pic,graphicx}
\usepackage{enumerate}
\usepackage{psfrag}
\usepackage{booktabs}
\usepackage{hyperref}      
\hypersetup{               
   pdftitle={mavisSPIE2020},    
   pdfauthor={GuidoAgapito},  
   colorlinks=true,      
   linkcolor=black,      
   anchorcolor=black,    
   citecolor=black,      
   filecolor=black,      
   menucolor=black,      
   pagecolor=black,      
   urlcolor=blue,         
   linktocpage=true
}

\usepackage{color, colortbl}
\definecolor{LightGray}{gray}{0.9}
\usepackage{diagbox}
\usepackage{array}
\newcolumntype{C}[1]{>{\centering\let\newline\\\arraybackslash\hspace{0pt}}m{#1}}

\title{MAVIS: System modelling and performance prediction} 

\author[a,c]{Guido Agapito}
\author[b,c]{Daniele Vassallo}
\author[a,c]{Cedric Plantet}
\author[b,c]{Valentina Viotto}
\author[a,c]{Enrico Pinna}
\author[d]{Benoit Neichel}
\author[d]{Thierry Fusco}
\author[e]{Francois Rigaut}
\affil[a]{INAF Osservatorio Astrofisico di Arcetri, L. Enrico Fermi 5, 50125 Firenze, Italy}
\affil[b]{INAF Osservatorio Astronomico di Padova, Vicolo dell’Osservatorio 5, 35122, Padova, Italy}
\affil[c]{ADaptive Optics National laboratory in Italy (ADONI)}
\affil[d]{Laboratoire d'Astrophysique de Marseille, 38 Rue Frédéric Joliot Curie, 13013 Marseille, France}
\affil[e]{AAO - Stromlo, RSAA, Australian National University, Cotter Road, Weston, ACT2600, Australia}

\authorinfo{Further author information:\\Guido Agapito, \Letter
 \hspace{0.5ex} guido.agapito@inaf.it\\}

\begin{document} 

  \maketitle

  \begin{abstract}
    The MCAO Assisted Visible Imager and Spectrograph (MAVIS) Adaptive Optics Module has very demanding goals to support science in the optical: providing 15\% SR in V band on a large FoV of 30arcsec diameter in standard atmospheric conditions at Paranal. It will be able to work in closed loop on up to three natural guide stars down to H=19, providing a sky coverage larger than 50\% in the south galactic pole. Such goals and the exploration of a large MCAO system parameters space have required a combination of analytical and end-to-end simulations to assess performance, sky coverage and drive the design. In this work we report baseline performance, statistical sky coverage and parameters sensitivity analysis done in the phase-A instrument study.
  \end{abstract}


%

\section{INTRODUCTION}
\label{sec:intro}

Numerical simulations have been used to support system design in selecting the baseline configuration of the AO module of MAVIS\cite{francois2020mavis} with main focus on the fulfilment of the Top Level Requirements (TLRs).
These requirements demand that the system reaches 10\% SR in V band on a FoV of 30arcsec diameter (15\% goal) in standard atmospheric conditions at Paranal with bright Natural Guide Stars ($\mathrm{m_J}$=8) and provides a sky coverage larger than 50\% in the south galactic pole guaranteeing an ensquared energy of at least 15\% on 50mas side in V band (in the same standard atmospheric conditions).

Most of the work has been done through, in particular, sensitivity analysis to different parameters, error terms, with a few different tools.
These tools that have also been used to compute synthetic PSFs to support the design of the MAVIS imager and spectrograph modules\cite{simon2020mavis,monty2020imager}, are presented in the next section, \ref{sec:tools}.
Then, the article continues with the description of the parameters of the baseline configuration of the MAVIS AO module\cite{valentina2020aom} in Sec. \ref{sec:params}, its error budget in Sec. \ref{sec:error}, and, finally, in Sec. \ref{sec:sens}, we briefly describe the sensitivity analyses done.

\section{SIMULATION TOOLS}\label{sec:tools}
To evaluate MAVIS AO module performance and to perform a comprehensive sensitivity analysis we used two main tools: the end-to-end INAF’s AO simulations tool PASSATA \cite{doi:10.1117/12.2233963} and a faster, Fourier-based tool, which is mostly based on the article by Neichel et al. 2009\cite{neichel2009}.
In this section we focus on a few important features of both of them. 


The PASSATA simulates all the processes that take place during the operation of an AO system, from input disturbance generation and propagation down to wavefront correction via the deformable mirrors (DMs).
Control implemented in end-to-end simulations is described in Ref.~\citenum{Busoni2019} and it is based on tomographic reconstruction and DM projection\cite{Fusco:01}, pseudo-open loop control (POLC)~\cite{10.1117/12.506580} 
and split tomography\cite{Gilles:08}.

In the next sub-sections we examine a few aspects of the simulation tools that may interest the reader: the generalized fitting error that is used to optimize the altitude and the pitch of the post focal DMs, the LGS sodium elongation that is required to account for the truncation effects and for the correct noise in the LGS sub-apertures, and the sky coverage estimation that is fundamental for the estimation of the statistics of the performance of the system.

\subsection{Generalized fitting}
Once we have the tomographic reconstruction of the turbulence, we need to apply a correction with the DMs.
This is done by minimizing the difference between the projection of the phase from the reconstructed layers and the projection of the phase from the DMs.
The error coming from this is called generalized fitting and it is studied to optimize the altitude and the pitch of the DMs.

We use a specific tool based on the approach described in Ref. \citeonline{Rigaut2000}.
Thanks to its low computation power requirements it is useful to evaluate a large combination of parameters as shown in Sec. \ref{sec:sens}.
We cross-checked the results of this tool with both PASSATA and Fourier\footnote{in these cases we compute generalized fitting error as the difference between a simulation with the actual DMs and one with a large number of DMs (equal to the number of reconstructed layers).} and we got good agreement.\\

\subsection{LGS sodium elongation}
PASSATA computes the LGS spot for each sub-aperture as a convolution of the PSF of a point source affected by the residual turbulence in conic propagation and the extended spots given by the laser geometry. The LGS spots modelling takes into account the broadening effect due to upward propagation (FWHM=1.2 arcsec), the geometry of the Shack-Hartmann WFS, of the launcher and of the sodium profile.
This convolution is done at high resolution with elements size of $0.5\lambda_{\mathrm{LGS}}/d_{\mathrm{sub-ap.}}$ and then the correct sampling of the detector is obtained with a down-scaling of the arrays.

\subsection{Sky coverage}
\label{sec:skycov}
While the high-order (modes above tip/tilt) residuals can be assessed independently from the observed FoV, the jitter residual has to be computed statistically, assuming a certain distribution of the stars. The sky coverage analysis is mainly performed with a code developed at Laboratoire d’Astrophysique de Marseille (LAM), with some cross-checks with a similar code developed at INAF/Arcetri. The pipeline consists of the following steps:
\begin{itemize}
  \item{\textbf{Asterism catalog generation.}} The source we use for stars statistics is the Besan\c{c}on galaxy model\cite{1986A&A...157...71R}.
  We get from the model a list of stars in a $3^{\circ}\times3^{\circ}$ field at the South Galactic Pole. These stars are placed with a uniform probability. We then generate a series of pointing coordinates and we register all 3-star asterisms at each pointing. Fields with only 1 or 2 stars are also registered.\\
  \item{\textbf{Low-order residual computation.}} For each registered asterism, we compute the LO residual (tip/tilt) in a given direction as follows:
  \begin{equation}
      \sigma_{LO} = \sqrt{\sigma_{vib}^2+\sigma_{tomo}^2+\sigma_{noise}^2}
  \end{equation}
  where $\sigma_{vib}^2$ is the tip/tilt Mean Square Error (MSE) due to vibrations (computed with formulas from Ref. \citeonline{10.1117/12.2232918}), $\sigma_{tomo}^2$ is the pure tomographic error\cite{Sasiela:94} and $\sigma_{noise}^2$ is the WFS noise term.
  
  \item{\textbf{Jitter in the scientific FoV}}. We compute the jitter in 5 directions in the scientific FoV: on axis and
  on 4 positions, close to the corners of the imager FoV, at 20" off axis.
  The average jitter in the FoV is simply found by averaging these 5 residuals.
  
  \item{\textbf{Sky coverage computation}}. The selected asterism for each pointing is the one giving the lowest average jitter in the scientific FoV. We then compute the sky coverage as the ratio between the number of fields giving a jitter less or equal to a given value and the total number of fields:
  \begin{equation}
    SC(x) = \frac{\sum_i P_i(\mathrm{Jitter} \leq x)}{N_{\mathrm{fields}}}
  \end{equation}
  with $P_i(\mathrm{Jitter} \leq x) = 1$ if the i-th field has an asterism giving a jitter less than $x$ and $P_i(\mathrm{Jitter} \leq x) = 0$ otherwise.
\end{itemize}

\section{Baseline configuration}\label{sec:params}

\begin{figure}[htp]
    \centering
    \includegraphics[width=14cm]{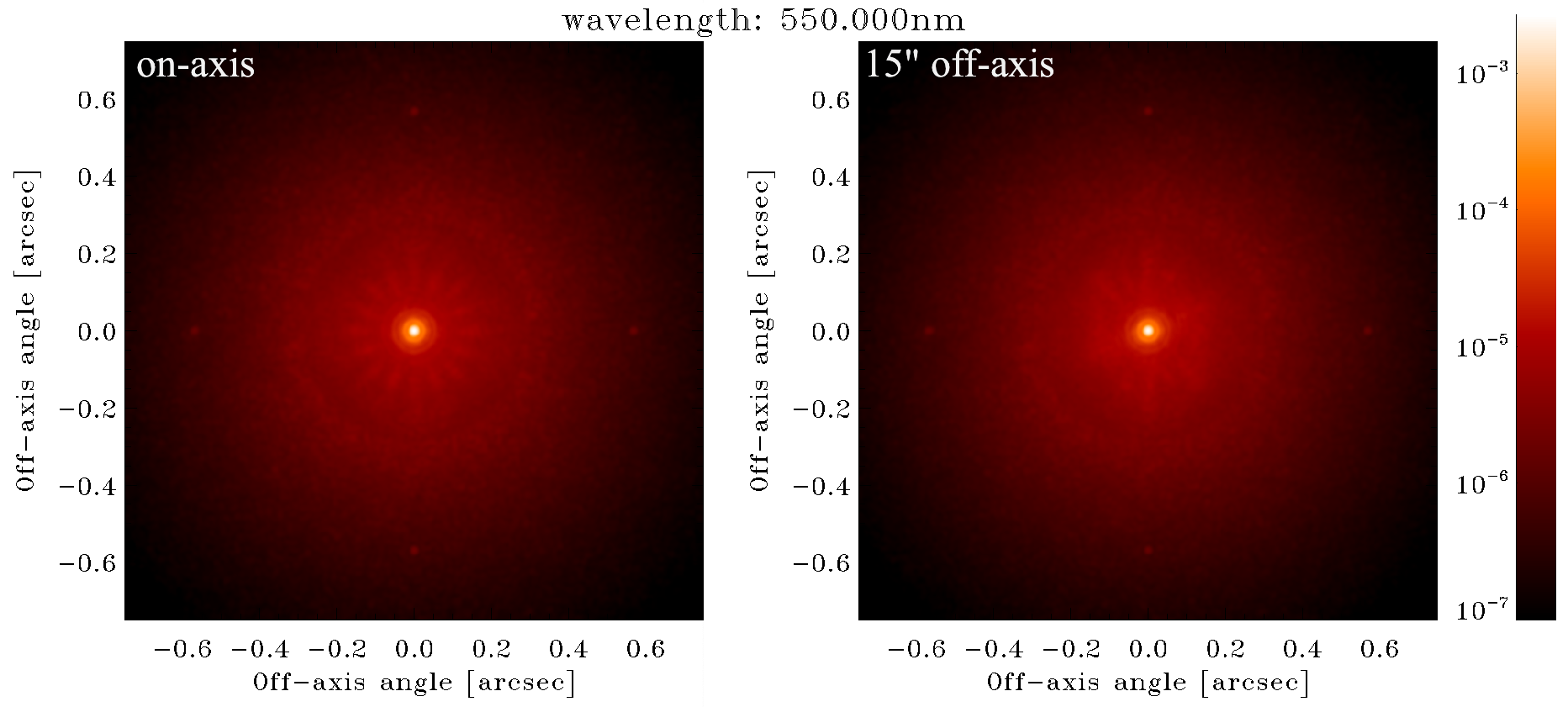}
    \caption{V band PSFs for the baseline configuration from an end-to-end simulation: left, on-axis one, right, 15arcsec off-axis one.}
    \label{fig:PSF}
\end{figure}

In this section we report the parameters used in the baseline configuration. They are summarized in Tab. \ref{tab:params}.
This configuration was chosen not only considering simulations results coming from the work reported here, but also considering all the aspects of the design of MAVIS\cite{francois2020mavis,valentina2020aom}. 
In Sec. \ref{sec:sens} some of these parameters were changed to explore all the selected different configurations. For example, for the study of the optimal LGS asterism with Fourier-based tool we used a set of Stereo-SCIDAR Cn$^{2}$ profiles\cite{10.1093/mnras/sty1070}.
In this baseline configuration we consider the pseudo open loop control with split tomography as described in Ref. \citeonline{Busoni2019}, but other simulation work focused on predictive control and learn and apply approach in the context of the AO module of MAVIS is reported in Ref. \citeonline{jesse2020l&a} and \citeonline{hang2020l&a} respectively.
This baseline configuration gives an average residual error over the science FoV of 125nm that is less than the 132nm of the requirement (that corresponds to 10\% SR in V band).\\As an example, a couple of V band PSFs from end-to-end simulations are shown in Fig. \ref{fig:PSF}.

%
\begin{table}[ht]
\caption{Summary of the baseline MAVIS parameters (in simulation).}
\label{tab:params}
\begin{center}
\begin{small}
	\begin{tabular}{|l|c|}
		\hline
		\textbf{Parameter} & \textbf{value}\\
		\hline
		Telescope diameter &  8m\\
		Central obstruction & 1.28m\\
		Pupil mask & round, no spiders\\
		Zenith angle & 30deg\\
		Science FoV (diameter) & 30\\
		Technical FoV (diameter) & 120\\
		\hline
		Atmospheric turbulence & 1 profile with 10 layers (Paranal median)\\
		$r_0$ & 0.126m\\
		$L_0$ & 25m\\
	    \hline
	    NGS full throughput & 0.28\\
		LGS full throughput (with laser splitting) & 0.184\\
	    \hline
		Ground DM & conjugated at 0m with $\sim$0.22m pitch\\
		Post focal DM no 1 & conjugated at 6000m with $\sim$0.25m pitch\\
		Post focal DM no 2 & conjugated at 13500m with $\sim$0.32m pitch\\ 
		\hline
		NGS WFS number & 3\\
		NGS WFS off-axis angle (good asterism) & 20 arcsec\\
		NGS WFS nSA & 1\\
		NGS WFS FoV & 3 arcsec\\
		NGS WFS pixel pitch & 30 mas\\
		NGS WFS detector RON & 0.5 $\mathrm{e^-/pixel/frame}$\\
		NGS WFS detector dark current & 20 $\mathrm{e^-/pixel/s}$\\
		NGS (H band) sky background & 2100 $\mathrm{e^-/m^2/arcsec^2/s}$ (no moon)\\
		NGS flux (full aperture, good asterism) & 1000 ph/ms\\
		\hline
		LGS WFS number & 8\\
		LGS WFS off-axis angle & 17.5 arcsec\\
		LGS WFS nSA & 40x40\\
		NGS WFS FoV & 5.0 arcsec\\
		LGS WFS pixel pitch & 0.866 mas\\
		LGS WFS detector RON & 0.2 $\mathrm{e^-/pixel/frame}$\\
		LGS launcher number & 4\\
		LGS launcher off-axis distance & 5.5 m\\
		LGS flux per sub-aperture (0.04 m$^2$) & 75 ph/ms\\
		\hline
		Sodium profile & ``multi peak''\cite{2014A&A...565A.102P}\\
		\hline
		Control & POLC with split tomography\cite{Busoni2019}\\
		Reconstruction layers altitude & [0, 0.33, 0.65, 1.25, 2.50, 5.00, 8.75, 12.50, 16.25] km \\
		Reconstruction layers pitch & [0.20, 0.20, 0.20, 0.20, 0.20, 0.22, 0.26, 0.30, 0.35] m\\
		Centroiding algorithm & CoG (LGS), weighted CoG (NGS)\\
		Framerate & 1000Hz\\
		Total delay & 2ms\\
		Integrator gains & 0.3-0.4 (LGS), 0.5 (NGS)\\
		\hline
		Extra error term & 28nm\cite{valentina2020aom}\\
		\hline
		\multicolumn{2}{c}{\scriptsize Note: NGS is Natural Guide Star, LGS is Laser Guide Star, DM is Deformable Mirror, WFS is Wavefront Sensor,}\\
		\multicolumn{2}{c}{\scriptsize SA is Sub-Aperture, FoV is Field of View, RON is Read-Out Noise, POLC is pseudo-open loop control and}\\\multicolumn{2}{c}{\scriptsize CoG is Center of Gravity.}\\
	\end{tabular}
\end{small}
\end{center}
\end{table}
\section{ERROR BUDGET}\label{sec:error}
We present in Table \ref{tab:budget.HO} the residual WFE breakdown from end-to-end simulations in the baseline configuration for both high and low orders. We quantified each error source individually. The sum of individual HO terms ($\sim$91 nm, summing up generalized fitting, tomographic, measurement noise, temporal and aliasing errors) is in good agreement with the HO residual of an end-to-end simulation in which all these error sources are included ($\sim$92 nm). The difference is +14 nm and could be due to the assumptions used to compute the error terms individually. In particular, it is difficult to fully isolate each of them and the computation assumes that they are independent, which is not exactly the case. This argument can explain why the root sum square of all individual terms is different than the end-to-end simulation result. Finally, we are assuming bright NGSs (m$_J$=8) for low orders sensing. In medium flux regime (m$_J$=15), $\sim$25 nm of measurement noise error have to be added. 

\begin{table}[h]
  \caption{Breakdown of MAVIS AO residual wavefront error.}
  \label{tab:budget.HO} 
  \smallskip
  \begin{minipage}{.5\linewidth}
    \centering
    \begin{tabular}{| l | c |}
      \hline
      \multicolumn{2}{| l |}{\textbf{High Orders}}\\\hline
      Error term & Error [nm]\\\hline
      High-frequency fitting error & 65.3\\ \hline
      Generalized fitting error & 30.1\\ \hline
      Tomographic error & 47.6\\ \hline
      Measurement noise error & 40.6\\ \hline
      Temporal error & 34.7\\ \hline
      Aliasing error & 40.1\\ \hline
      Sodium elongation/truncation & 23.6\\ \hline
      LGS jitter & 6.1\\ \hline
    \end{tabular}\smallskip
  \end{minipage}
  \begin{minipage}{.5\linewidth}
    \centering
    \begin{tabular}{| l | c |}
      \hline
      \multicolumn{2}{| l |}{\textbf{Low Orders}}\\\hline
      Error term & Error [nm]\\\hline
      Tomographic error & 38.0\\ \hline
      Measurement noise error & $\sim$0\\ \hline
      Temporal error & 29.2\\ \hline
      Wind shake/vibrations & 5.5\\ \hline
    \end{tabular}\smallskip
  \end{minipage}
\end{table}


\section{SENSITIVITY ANALYSIS}\label{sec:sens}


\paragraph{Asterism}
As a preliminary step, we explored the system sensitivity to the geometry of the LGS asterism, both in terms of number of stars and asterism radius. Since we included generalized fitting error in these simulations, we also considered DM pitches as a third parameter to explore. The purpose of this analysis is to make a first guess of the most interesting system configurations and to start constricting the parameter space. As it concerns the DM pitches, not all the values we explored are even practically feasible.\\For each combination of the three parameters, we computed with Fourier the overall contribution of tomographic, full fitting (generalized+high-frequency) and noise errors, averaged over MAVIS field of view and over 243 Stereo-SCIDAR profiles\cite{10.1093/mnras/sty1070} corresponding to median seeing conditions (0.7” @z=$30^{\circ}$). Results are reported in Table \ref{tab:astPitch}.

\begin{table}[h]
  \centering
  \caption{Residual wavefront error with Fourier (limited to tomographic, fitting and noise error) for several combinations of number of guide stars, asterism radius and DM pitches.}
  \label{tab:astPitch} 
  \smallskip
  \begin{tabular}{| C{1.5cm} | C{1.6cm} | c | c | c |}
    \hline
    Asterism radius [arcsec] & DM pitches [cm] & \multicolumn{3}{ c |}{Residual [nm]}  \\ 
    \cline{3-5}
    && 4 LGS & 6 LGS & 8 LGS \\
    \hline
    17.5 & (30,40) & 111 & 98 & 96\\ \hline
    17.5 & (30,30,40) & 103 & 87 & 85\\ \hline
    17.5 & (20,20,20) & 94 & 75 & 71\\ \hline
    25.0 & (30,40) & 128 & 104 & 100\\ \hline
    25.0 & (30,30,40) & 122 & 95 & 90\\ \hline
    25.0 & (20,20,20) & 115 & 85 & 78\\ \hline
    30.0 & (30,40) & 140 & 110 & 103\\ \hline
    30.0 & (30,30,40) & 135 & 102 & 94\\ \hline
    30.0 & (20,20,20) & 130 & 93 & 84\\ \hline
  \end{tabular}
\end{table}

\begin{figure}[htp]
    \centering
    \includegraphics[width=8cm]{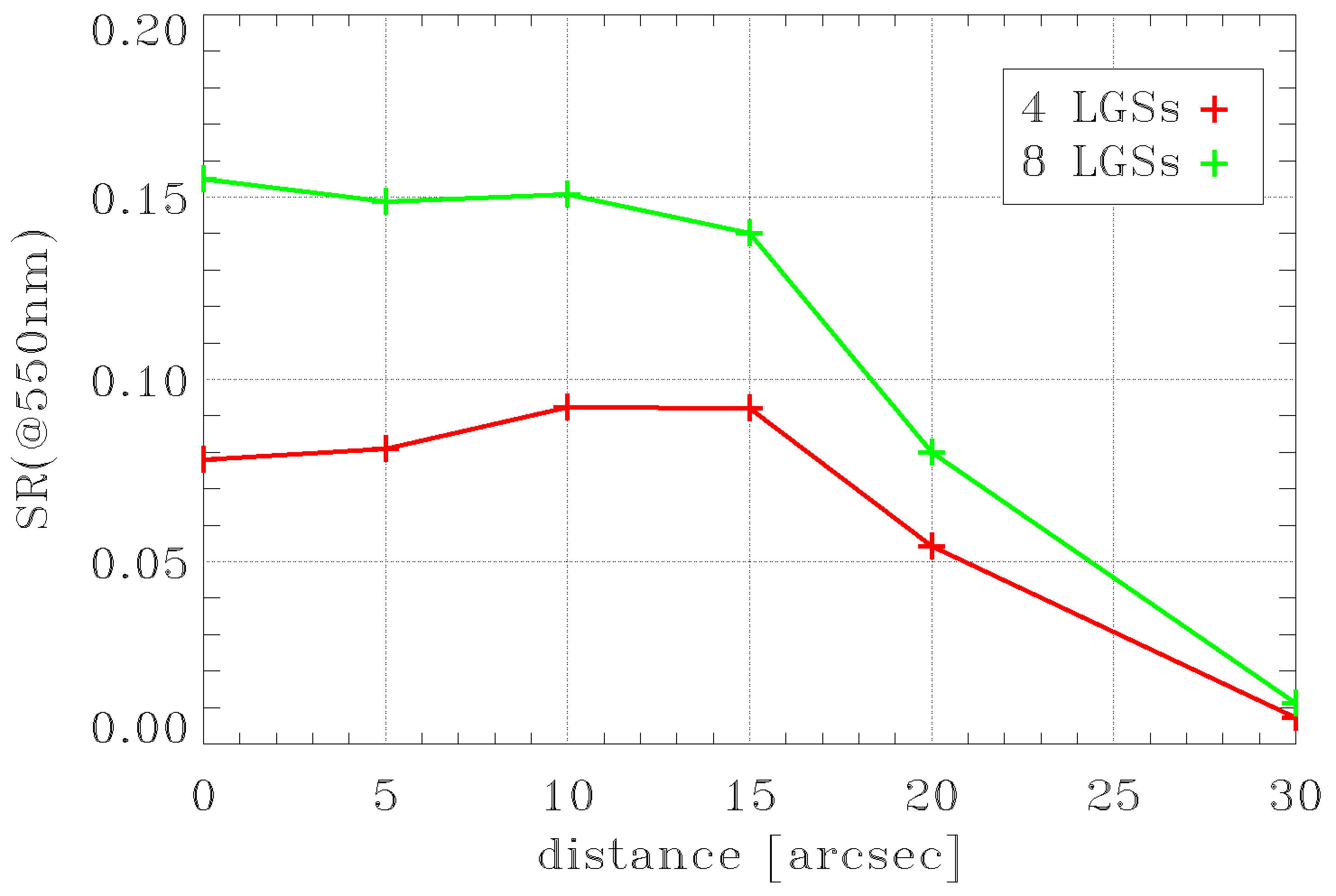}
    \caption{V-band SR (from end-to-end) versus off-axis angle for 4 and 8 LGSs, 17.5" radius and baseline configuration (see Tab.\ref{tab:params}).}
    \label{fig:SR4vs8}
\end{figure}

One of the most interesting comparisons among the ones explored with Fourier is between 4 and 8 LGSs for the smallest asterism radius.
We ran end-to-end simulations over these two configurations with the baseline configuration (see Tab.\ref{tab:params}) and we found that only 8 LGSs are able to meet the requirement of 10\% SR in V-band with the ESO-TLR atmospheric profile (Figure \ref{fig:SR4vs8}).
It is also important to point out that performance from these end-to-end simulations cannot be directly compared with Fourier because we are using different Cn$^{2}$ profiles and, moreover, our Fourier simulations only account for a few error terms (tomographic, fitting and noise) out of the whole budget.

\begin{figure}[htp]
    \centering
    \includegraphics[width=8cm]{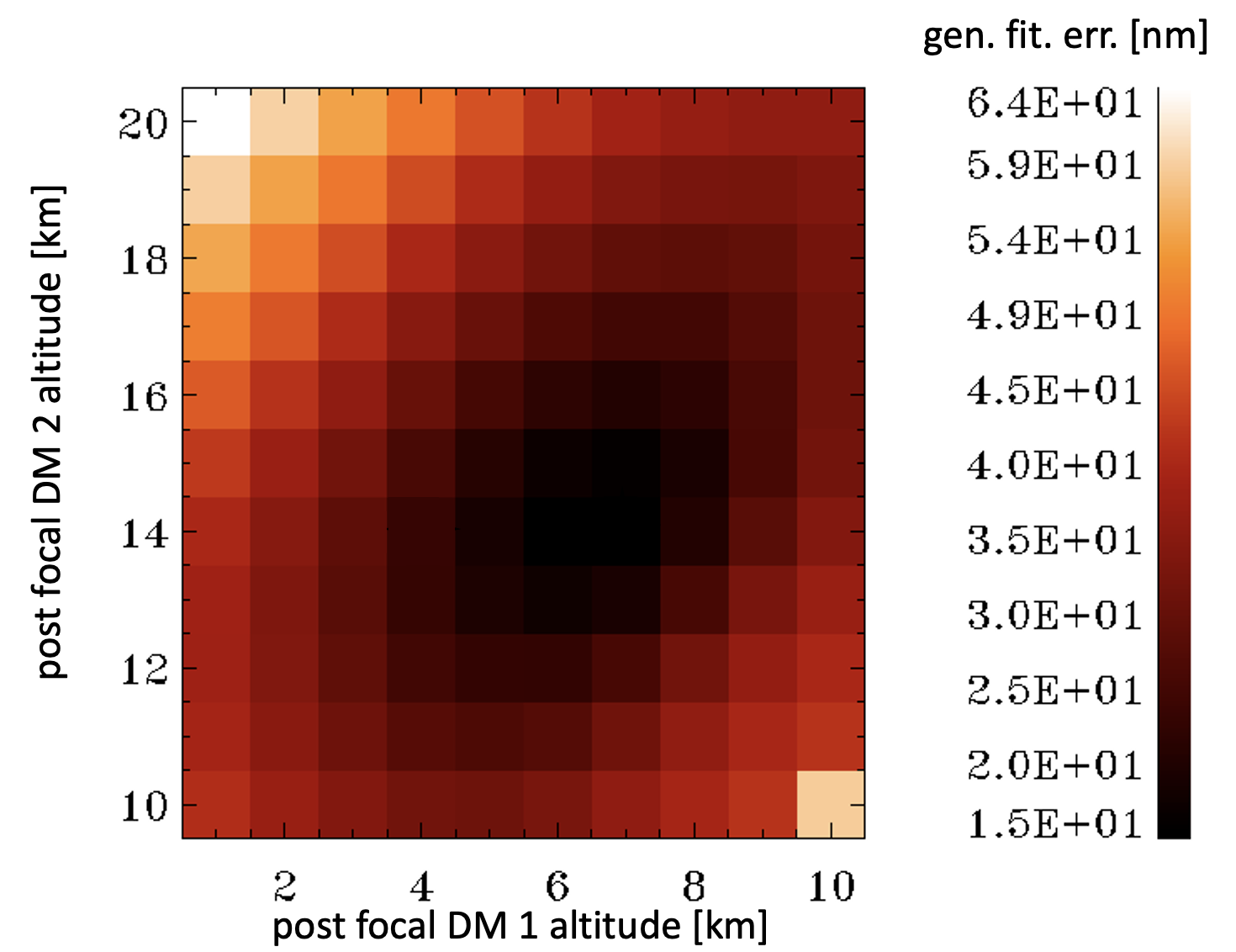}
    \caption{Generalized fitting error (with Fourier) as a function of the altitude of post-focal DMs.}
    \label{fig:genFit}
\end{figure}

\paragraph{Generalized fitting}
We calculated the generalized fitting error for several altitudes of the post-focal DMs using Fourier. The analysis shows that good ranges are 5-7km and 13-15km (figure \ref{fig:genFit}). 

\paragraph{Number of sub-apertures}
We report in Table \ref{tab:subap} the results of the sensitivity analysis on the number of sub-apertures of the LGS WFSs. In this analysis (performed with PASSATA) we also explored three different LGS flux levels. Noise covariance used to compute the tomographic reconstructor has been optimized iteratively for each combination of the two parameters. We found very similar results for the $36\times36$ and $40\times40$ cases in the flux regime we are mainly interested in. Further simulation work is needed in the next phase to validate these preliminary conclusions.  
\smallskip
\noindent
\begin{table}[!h]
  \centering
  \caption{Relative on-axis residual wavefront error in nm (with PASSATA) as a function of the number of sub-apertures (Nsa) of the LGS WFSs and of the LGS flux (in ph/ms/subap on a $40\times40$ WFS). The reference configuration has a 0 relative error.}
  \label{tab:subap} 
  \smallskip
  \begin{tabular}{|c||*{3}{c|}}\hline
     \backslashbox[20mm]{Flux}{Nsa}
     &\makebox[3em]{32}&\makebox[3em]{36}&\makebox[3em]{40}\\\hline\hline
     37.5  & +43.4 & +37.3 & +40.1\\\hline
     75   & +17.8 & -6.0 & 0 \\\hline
     150 & -24.1 & -29.1 & -29.6\\\hline
  \end{tabular}
\end{table}

\paragraph{Temporal error}
We ran LGSs-only simulations in PASSATA to quantify the sensitivity to average wind speed and AO loop delay. Results are reported in Table \ref{tab:tempErr}.
\smallskip
\noindent
\begin{table}[h]
  \centering
  \caption{Relative on-axis residual wavefront error in nm (with PASSATA) as a function of AO loop delay (in frames) and average wind speed (in m/s). The reference configuration has a 0 relative error.}
  \label{tab:tempErr} 
  \smallskip
  \begin{tabular}{|l||*{4}{c|}}\hline
     \backslashbox[32mm]{delay}{wind speed}
     &\makebox[3em]{1.0}&\makebox[3em]{5.1}&\makebox[3em]{10.2}
     &\makebox[3em]{20.4}\\\hline\hline
     2.0 & -36.4 & -31.9 & 0 & +67.6\\\hline
     2.5 & -34.6 & -29.5 & +17.2 & +73.5\\\hline
     3.0 & -19.4 & -0.5  & +36.7 & +88.0\\\hline
  \end{tabular}\smallskip
\end{table}

\begin{figure}[h!]
    \centering
    \includegraphics[width=8cm]{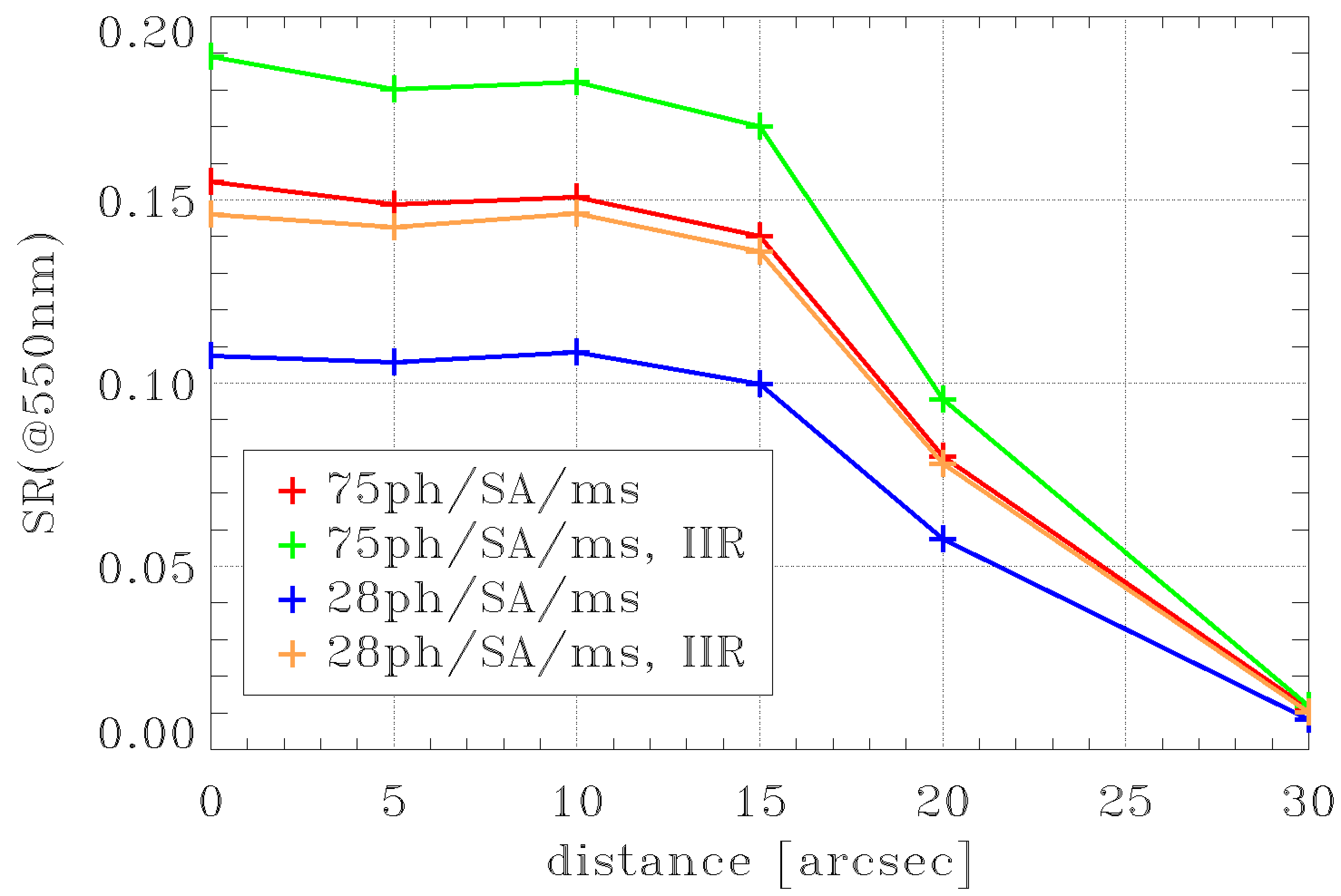}
    \caption{V-band SR (from end-to-end) in function of off-axis angle for standard flux condition and 90\% flux condition with integrator and infinite impulse response (IIR) filters (optimized as show in Sec.\ref{sec:sens}) controllers.}
    \label{fig:SRiir}
\end{figure}

\paragraph{Temporal control filters}
We analysed the possibility to use higher order filters (infinite impulse response one) instead of pure integrators to better shape closed loop transfer functions and improve performance thanks to a better rejection of temporal and noise errors, in particular with lower fluxes.
We have limited our analysis to the HO/LGSs control, but it could be extended to LO/NGSs too.
The optimization of this filter for a system with pseudo open loop control is not trivial because POLC introduces a multi-input multi-output dynamic control term given by the matrix $H$ described in Ref.~\citeonline{Busoni2019}.
Currently, we have no way to consider this term in the control filter optimization, and, so, we are limited to optimization tools designed for single conjugate AO systems\cite{AgapitoSPIE2012,10.1117/1.JATIS.5.4.049001}.
Our approach has been a trial-and-error one, where we optimize filters with different turbulence parameters and noise level and we run end-to-end simulation to evaluate their performance.
A comparison between integrator and the optimized filters based controller is presented in Fig.\ref{fig:SRiir}.
In particular, these filters are able to improve performance in low flux regime: considering a sodium return flux of 0.4e7 ph/s/m$^2$ (corresponding to 28 ph/ms/sa) the improvement is larger than 3\% V band SR with respect to a pure integrator controller, and gets very close to the same performance (difference is $\sim$1\% V band SR) of a pure integrator with a higher return flux of 1.07e7 ph/s/m$^2$ (75 ph/ms/sa). But even in the highest flux condition considered, 75 ph/ms/sa, they lead to an improvement of $\sim$ 3\% V band SR.

\paragraph{Seeing}
For this analysis, we ran a batch of end-to-end simulations changing the seeing parameter while keeping the same normalized Cn$^{2}$ profile. For simplicity, we ran simulations with LGSs only, so LGSs are used to measure tip/tilt too. Results show that residual variance follows approximately a $(r_{0a}/r_{0b})^{5/3}$ law. This is expected since, as can be seen in Ref. \citeonline{10.1117/12.321649}, fitting error and Shack-Hartmann WFS aliasing error follow a $(d/r_0)^{5/3}$ law and fitting error is the largest term in our AO error budget.


\paragraph{Spot elongation}
The LGS spot in a given sub-aperture is characterized by an elongation and orientation that both depend on the position of the sub-aperture itself with respect to the launcher. A possible configuration to maximize the available WFS FoV (that is square) is to rotate the WFS such that the most elongated spots fall on the diagonal of the sub-aperture (the $45^{\circ}$ configuration, hereafter). This is important because truncated spots (i.e. spots that are not fully imagined in the sub-apertures FoV) give a not null signal even with a flat wavefront.\\We ran a batch of simulations considering two different sodium profiles and two WFS orientations. Sodium profiles are the “multi-peak” and “very wide” ones from Ref. \citeonline{2014A&A...565A.102P}, which are fairly representative of a “good” and a “bad” case, respectively. WFS orientations are $0^{\circ}$ (non-optimized) and $45^{\circ}$ (optimized). Results are reported in Table \ref{tab:elong}.
\begin{table}
  \centering
  \caption{Average high orders (tip/tilt excluded) wavefront error in nm (with PASSATA) due to LGS spot elongation in science and technical FoV for a 0 and 45$^{\circ}$ configurations.}
  \label{tab:elong} 
  \smallskip
  \begin{tabular}{|l|c|c|c|c|}
    \hline
    \backslashbox[40mm]{FoV}{Na profile} & \multicolumn{2}{c|}{Very Wide} & \multicolumn{2}{c|}{Multi-Peak} \\ \hline
    conf. angle [deg] & $\; \; 0 \; \;$ & $45$ & $\; \; 0 \; \;$ & $45$\\ \hline
    Science (r$\leq$15") & $36$ & $30$ & $25$ & $24$\\
    Technical (15"$<$r$\leq$60") & $62$ & $54$ & $38$ & $38$\\
    \hline
  \end{tabular}
\end{table}

\begin{figure}[h]
  \centering
  \includegraphics[width=12cm]{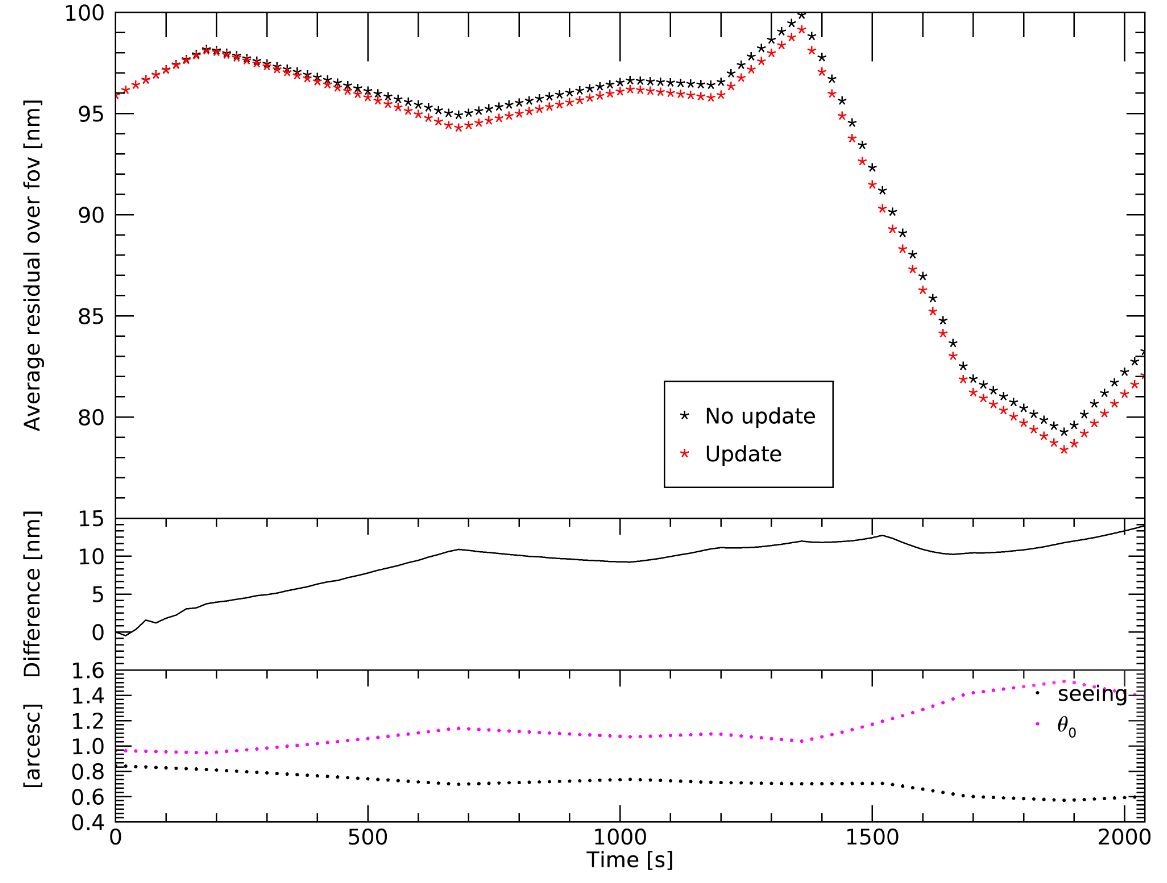}
  \caption{Top panel: absolute residual wavefront error with Fourier (averaged over the FoV) as a function of time, with and without updating the tomographic reconstructor. Middle panel: the difference between the two curves. Bottom panel: temporal trends of seeing and $\theta_0$.}
  \label{fig:cn2.update}
\end{figure}

\paragraph{Cn$^2$ update}
Tomographic reconstruction requires knowledge of the turbulence statistics. In simulations (both PASSATA and Fourier) we can use our direct knowledge of the input 3D atmospheric disturbance to build a "perfect" turbulence covariance matrix. If the Cn$^2$ profile varies during the night, the reconstructor should in principle be updated, otherwise performance might be affected. To investigate this effect, we selected 4 consecutive stereo-SCIDAR profiles\cite{10.1093/mnras/sty1070} (spanning half an hour in total) and we linearly interpolated them to get a time resolution of 20 seconds. For each time point we ran the Fourier code to compute the residual wavefront error, both with and without updating the reconstructor. Results are shown in Figure \ref{fig:cn2.update}. The loss in performance without the update is stable after 10 minutes between 10 and 15nm. We also ran end-to-end simulations in the same conditions and we found a good agreement with Fourier results. 

\subsection{Sky coverage}
Here we report the sensitivity analysis on Low Orders, i.e. on the correction given by the NGSs on tip/tilt and quadratic term of plate scale. The residual of this correction -the residual jitter- depends on the NGS asterism we have for a given observation. Following step-by-step the procedure outlined in Section \ref{sec:skycov}.
  
\begin{figure}[htp]
    \centering
    \includegraphics[width=11cm]{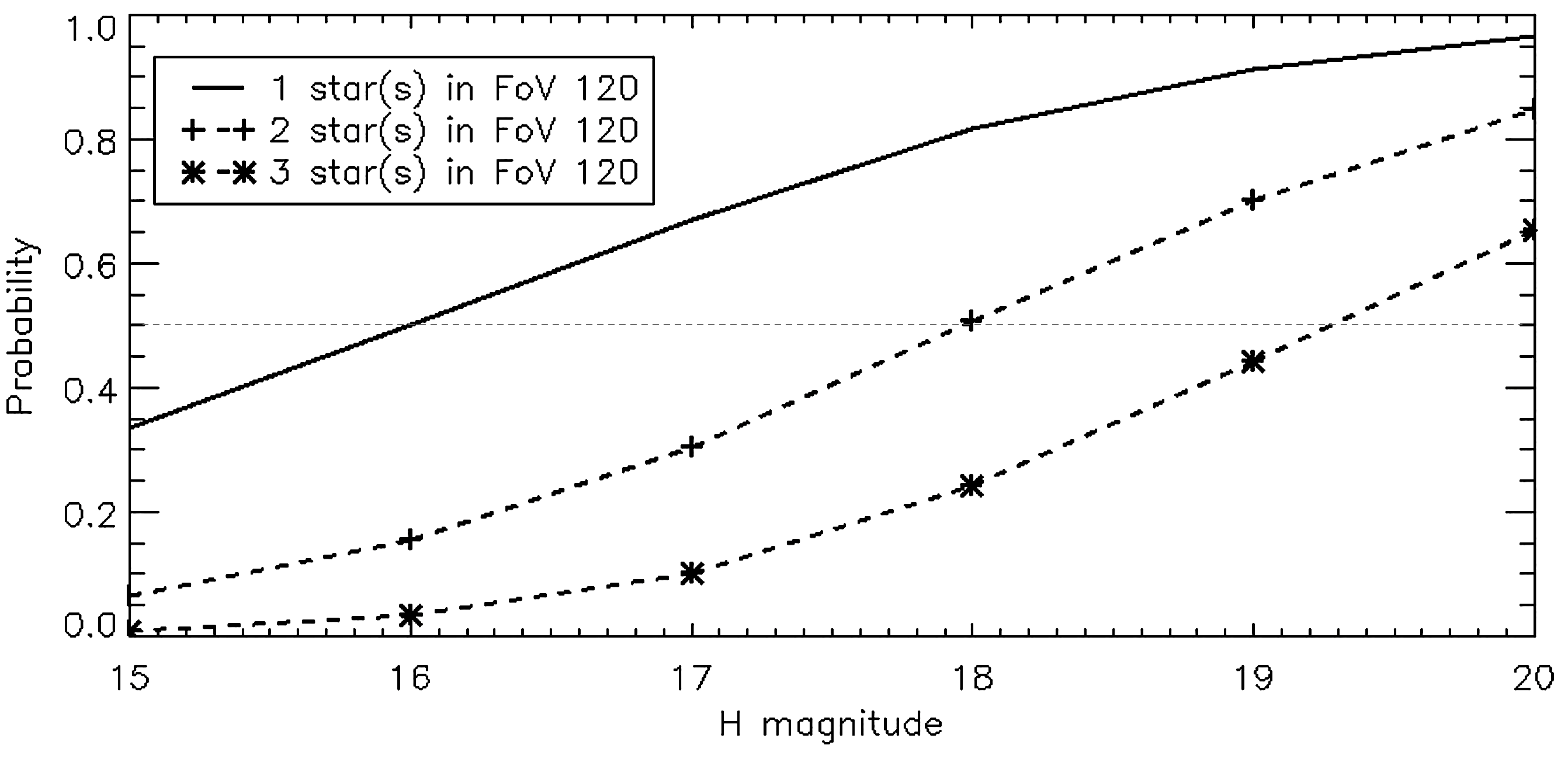}
    \caption{Probability of finding 1, 2 or 3 stars in MAVIS technical FoV when pointing towards the South Galactic Pole.}
    \label{fig:skycov.prob}
  \end{figure}

\begin{figure}[h!]
    \centering
     \subfigure[Jitter due to temporal error as a function of LO frame rate.\label{fig:skycov.temp}]
    {\includegraphics[width=0.42\columnwidth]{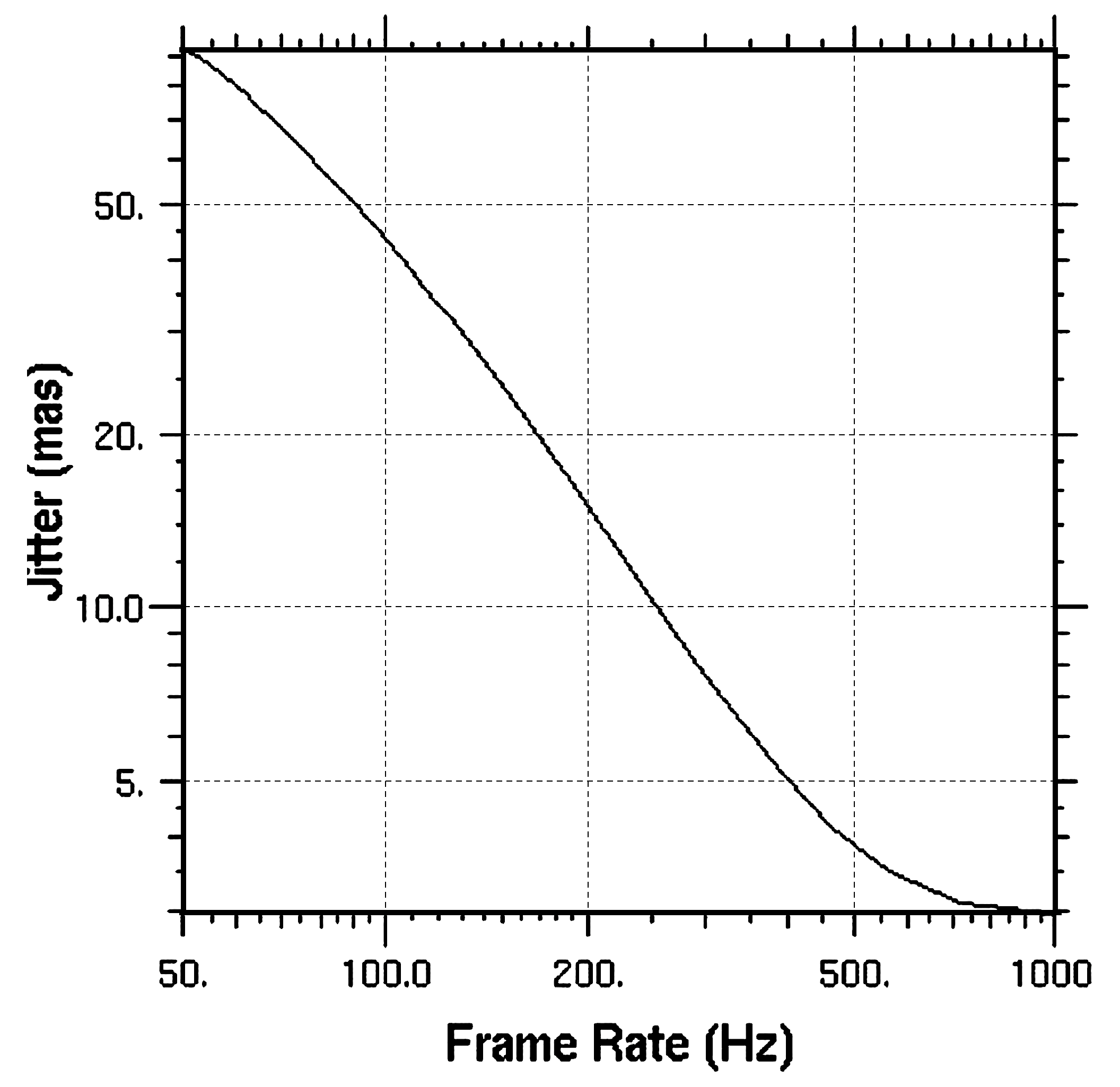}}
    \subfigure[Jitter due to measurement noise as a function of number of photons per sub-aperture per frame.\label{fig:superShAndRot}]
    {\includegraphics[width=0.40\columnwidth]{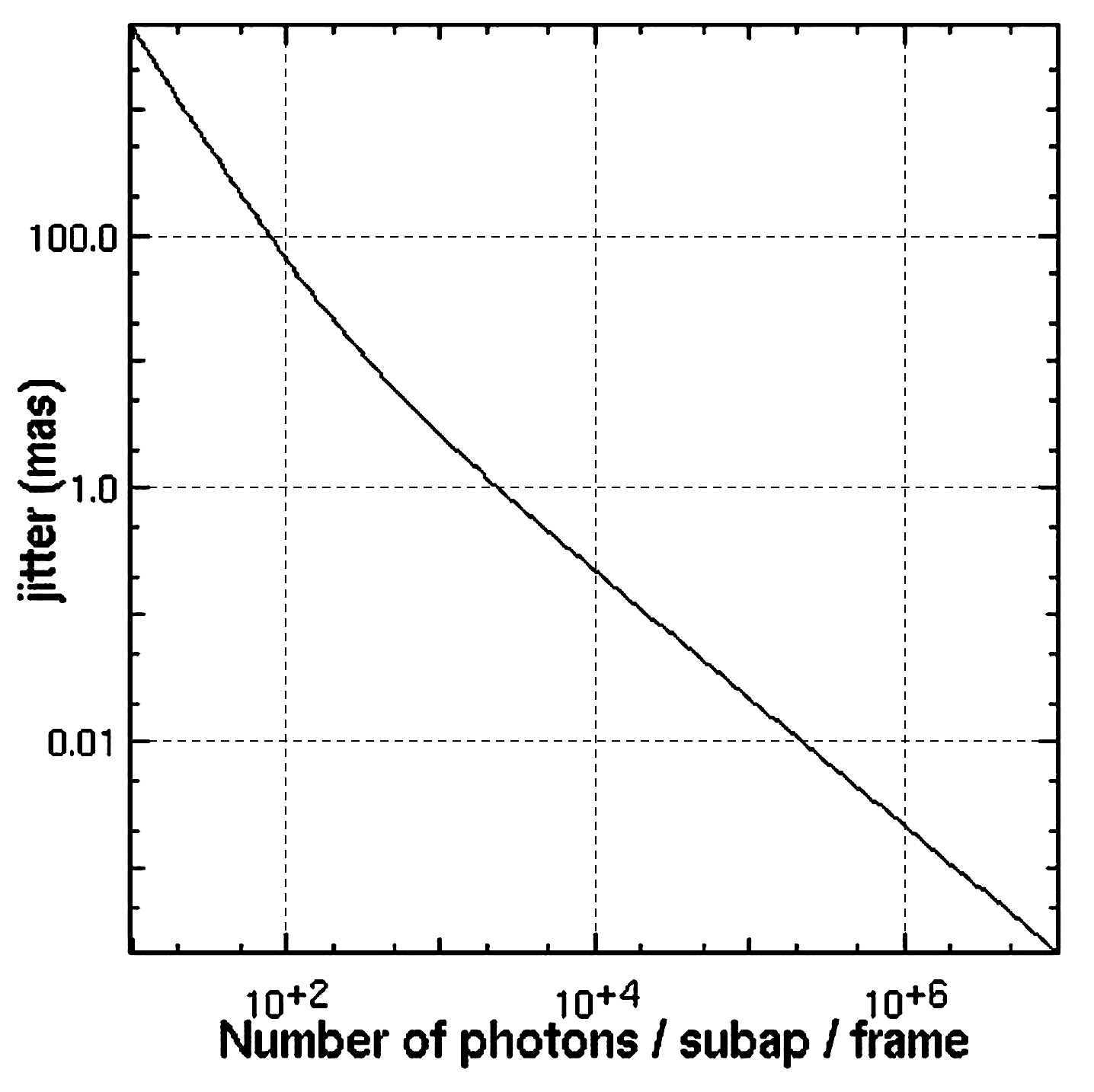}}
    \caption{Jitter error in mas.}\label{fig:super}
  \end{figure}

\begin{figure}[htp]
    \centering
    \includegraphics[width=11cm]{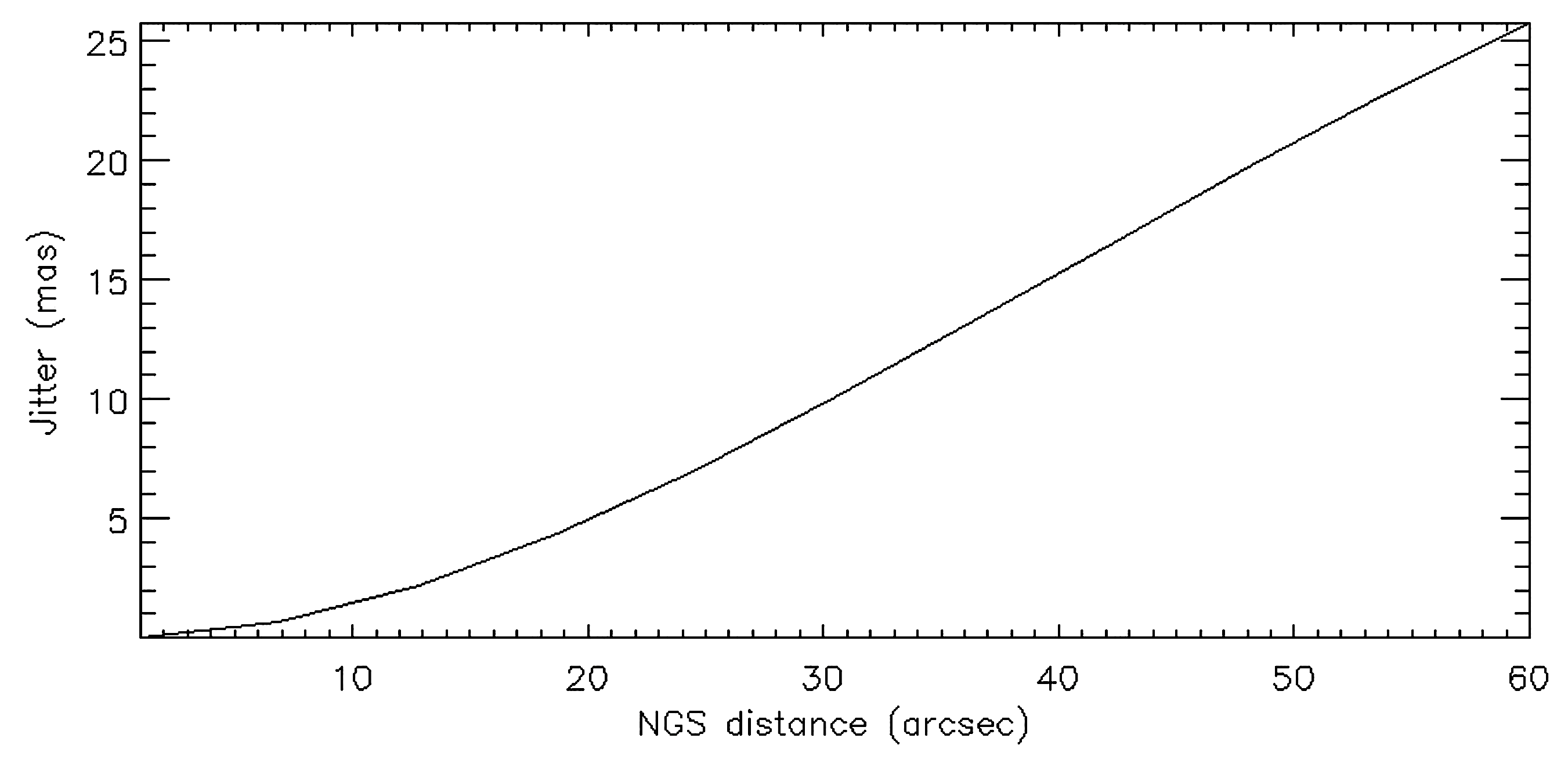}
    \caption{“Tomographic” Jitter (on-axis) as a function of the distance from the center.}
    \label{fig:skycov.tomo}
  \end{figure}

\begin{figure}[htp]
    \centering
    \includegraphics[width=14cm]{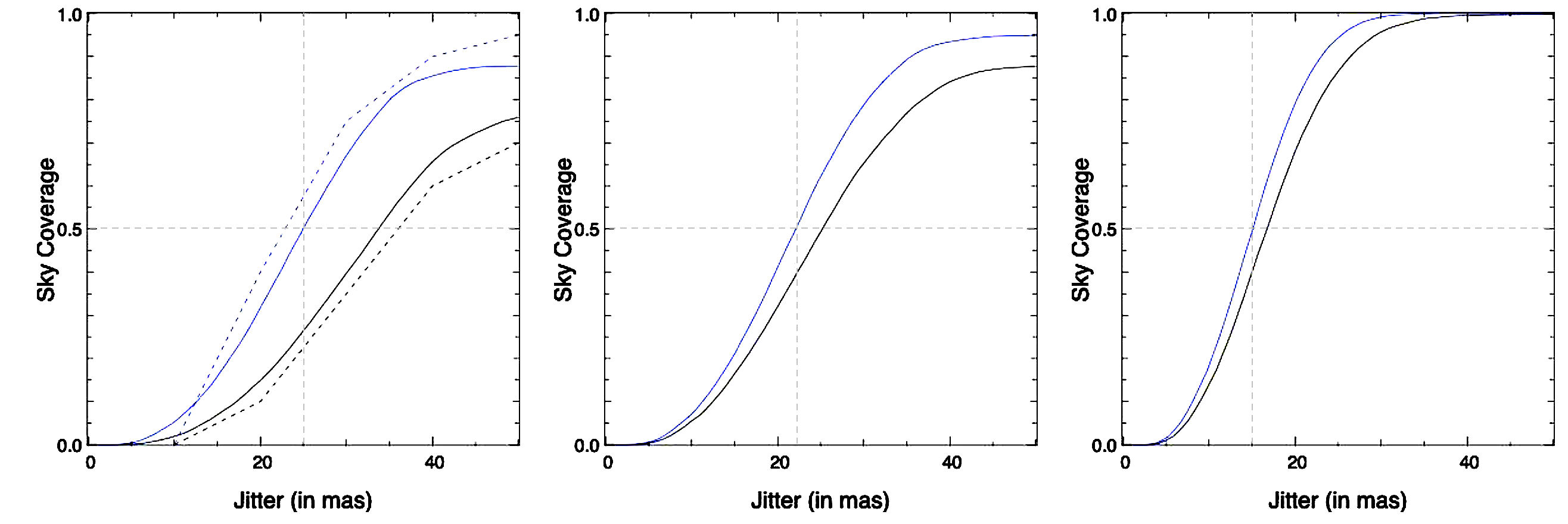}
    \caption{Sky coverage: left, for the galactic pole, middle, 60degrees above the galactic plane and, right, 30 degrees above the plane. Blue lines are for a 1$\times$1 WFS and black lines are for a 2$\times$2 WFS. For the galactic pole, the dashed plots are the results when using the Arcetri simulation tool.}
    \label{fig:skycov}
  \end{figure}

\paragraph{Asterism catalog generation.}
The statistics of the available stars in the technical FoV is plotted in Figure \ref{fig:skycov.prob}. No star of magnitude H $\leq20$ is available in about 5\% of the cases, while only one H $\leq20$ star is available in 15\% of the cases, excluding the possibility to correct quadratic plate scale modes. Then, the number of available H $\leq20$  stars is less than 3 in 35\% of the case. Finally, a star of magnitude H $\leq16$  is available in 50\% of the cases, providing enough flux to get one tip/tilt measurement at a high frame rate (for effective wind shake/vibrations compensation).
\paragraph{Low-order residual computation.} 
As detailed in Section \ref{sec:skycov}, the three dominant contributors to the overall residual jitter are temporal error, noise and tomographic error. The first is mostly dominated by windshake residuals and only depends on the NGS loop frequency. The residual vs framerate is plotted in Figure \ref{fig:skycov.temp}, for a global jitter that comprises both windshake/vibrations and atmosphere. Noise jitter versus flux is plotted in \ref{fig:superShAndRot}. We assumed for this calculation a 50\% SR, a noise level on the detector of 1 e- RMS, 
an overall transmission of 0.25 for the NGS WFSs and a bandpass including both H and J bands. In these conditions, a H=16 magnitude star would give $\sim$50 photons/frame/VLT pupil (at 500Hz). The conversion from photons to noise is based on noise propagation computations.\\Finally, Figure \ref{fig:skycov.tomo} shows the tomographic error (on-axis) for an equilateral constellation of 3 stars. The absolute amount of jitter due to tomography depends on the Cn$^{2}$ profile and the outer scale of the turbulence. 
  %
  \smallskip
%
\paragraph{Sky coverage computation.} 
In Table \ref{tab:skycov.results} we summarize the expected jitter for different sky coverage percentages at two galactic altitudes.
The sky coverage curves are shown in Fig. \ref{fig:skycov}.
We add the 2$\times$2 sub-apertures case to check the impact on jitter of having three of such sensors that are able to measure tip/tilt and give the truth sensing on focus and astigmatisms.
The main conclusion is that for 50\% Sky Coverage at the pole, we should expect a jitter of about 25mas. 
  \begin{table}[!ht]  
    \centering
    \caption{Expected jitter at South Pole and 30$^\circ$ galactic altitude as a function of sky coverage.}
    \label{tab:skycov.results} 
    \smallskip
    \begin{tabular}{| C{1.5cm} | c | c |}
      \hline
      Sky & \multicolumn{2}{ c |}{jitter [mas]}\\ 
      \cline{2-3}
      Coverage & Pole & 30$^\circ$\\
      \hline
      20\% & 16.5 & 10 \\ \hline
      50\% & 25   & 15 \\ \hline
      70\% & 30   & 18 \\ \hline
    \end{tabular}\smallskip
  \end{table}
%

\section{Conclusion and future work}
MAVIS design is still in the preliminary phase, but we already have a set of tools for system modelling and performance prediction.
We have shown how we verified that the baseline configuration is able to guarantee the fulfilment of the TLRs.

In the next design phase we will progress in the sensitivity study analysis and update the performance prediction with a more accurate model of the system (for example considering the influence functions of the deformable secondary mirror of the VLT\cite{10.1117/12.2057591} and the telescope spiders).
Moreover, we will revise our control scheme in coordination with the real time computer team\cite{damien2020rtc} to improve the final performance to be more robust to some critical aspect such sodium return flux.

\bibliography{biblio}
\bibliographystyle{spiebib}

\end{document}